\documentclass[prl,aps,twocolumn,groupedaddress,showpacs,amsmath,amssymb]{revtex4-1}
\usepackage{graphics}
\usepackage{epsfig}
\usepackage[usenames,dvipsnames]{color}
\usepackage{rotating}
\usepackage{times}

\newcommand{\MakeFootnoteRef}[1]{\newcounter{#1}\setcounter{#1}{\value{footnote}}}

\definecolor{MyOrange}{RGB}{226,116,5}
\definecolor{MyTurquoise}{RGB}{0,51,51}
\definecolor{MyPurple}{RGB}{102,0,81}

\begin{document}

\title{$S$-Wave Collisional Frequency Shift of a Fermion Clock}

\author{Eric L. Hazlett, Yi Zhang, Ronald W. Stites, Kurt Gibble and Kenneth M. O'Hara}

\affiliation{Department of Physics, The Pennsylvania State University,
University Park,\nolinebreak \,Pennsylvania 16802}

\begin{abstract}
We report an $s$-wave collisional frequency shift of an atomic clock based on fermions. In contrast to bosons, the fermion clock shift is insensitive to the population difference of the clock states, set by the first pulse area in Ramsey spectroscopy, $\theta_1$. The fermion shift instead depends strongly on the second pulse area $\theta_2$. It allows the shift to be canceled, nominally at $\theta_2 = \pi/2$, but correlations shift the null to slightly larger $\theta_2$. The shift applies to optical lattice clocks and increases with the spatial inhomogeneity of the clock excitation field, naturally large at optical frequencies. \vspace{-0.2in}
\end{abstract}

\pacs{06.30.Ft, 34.50.Cx, 06.20.-f, 37.10.Jk \vspace{-0.15in}}

\maketitle

At ultracold temperatures, atom-atom interactions can only occur through $s$-wave collisions.  While $s$-wave collisions are allowed for bosons and are the most important limitation to the the accuracy of clocks that establish international atomic time~\cite{Bize2012,Szymaniec2011}, they are forbidden for identical fermions by the Pauli exclusion principle.  Even when dephasing made fermions distinguishable, collision shifts were absent~\cite{Ketterle2003B}.  Thus,  ultracold fermions were thought to be immune to $s$-wave collisional frequency shifts ($s$CFS's) making them  ideally suited for precision metrology~\cite{Verhaar1995,Ketterle2003A,Katori2005,Inguscio2004,Romalis2007} and quantum memories~\cite{Hau2001,Kimble2005,Kuzmich2005,JianWei2012}. However, recent theoretical work predicted that fermions can have an $s$CFS because generally occurring inhomogeneities of excitation fields makes particles distinguishable~\cite{Gibble2009,Rey2009,Pethick2010}.

\vspace{-0.025in}

\begin{figure}[h]
\vspace{0.025in}
\includegraphics[width=0.9\columnwidth,angle=0,clip=true]{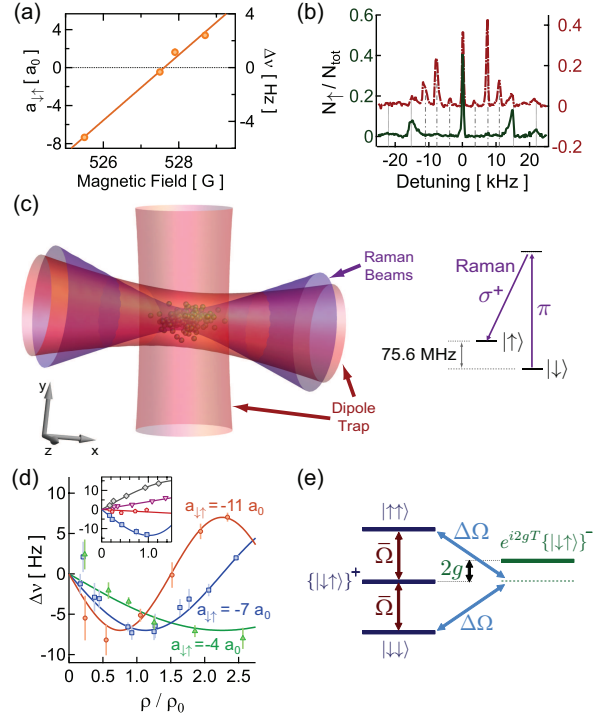}
\vspace{-0.0875in}
\caption{(color online). (a) The $s$-wave scattering length $a_{\downarrow \uparrow}$ in $^6$Li tunes linearly through zero at a bias field of 527.5~G, giving an unambiguous signature to the $s$-wave collisional frequency shift. (b) Rabi spectra in the resolved sideband limit for Raman beams centered on (solid) or offset from (dashed) the cloud center.  (c) Two copropagating Raman beams, focused to a waist comparable to the size of the atomic cloud, provide a spatially inhomogeneous excitation.  (d) Measured frequency shifts versus density for several $a_{\downarrow \uparrow}$, demonstrating a sinusoidal dependence characteristic of spin-waves. (Inset) The dependence at low-density gives the linear shift, shown for the four values of $a_{\downarrow \uparrow}$ in (a).  (e) Singlet and triplet states for two fermions in different trap states. A spatial inhomogeneity $\Delta \Omega$ couples triplet states to the singlet state, which has collisional interactions $g \propto a_{\downarrow \uparrow}$.
}
\vspace{-0.225in}
\label{Fig1}
\end{figure}

Here we experimentally observe an $s$-wave collisional frequency shift of an atomic clock based on a thermal gas of ultracold fermions.  Ramsey spectroscopy clearly distinguishes the novel behaviors of the $s$CFS, via specific dependences on the first and second Ramsey pulse areas, $\theta_1$ and $\theta_2$. We demonstrate that the shift is insensitive to $\theta_1$ and thereby the difference of the spin populations~\cite{Gibble2009}, in stark contrast with the shifts for bosons and the often-used mean-field expression.  Instead, the fermion $s$CFS depends strongly on $\theta_2$, which reads out the interaction induced phase shifts of each atom.  The shift is canceled if the atoms' phases are detected, on average, with equal sensitivity~\cite{Gibble2009}.  Interestingly, we show that correlations in the sample perturb the null of the $s$CFS to $\theta_2$ slightly greater than $\pi/2$. We explicitly see that the $s$CFS increases as expected with the inhomogeneity of the clock field, which we characterize independently. The fermion $s$CFS we observe in the resolved sideband regime is exactly analogous to those of optical lattice clocks~\cite{Katori2005}, for which the spatial field inhomogeneity is naturally large at optical frequencies. Recently, the fermion $s$CFS was simulated using an $^{87}$Rb Bose gas ~\cite{Rosenbusch2012}. They worked, in contrast, with unresolved trap sidebands to directly excite a spin-wave, and observed the predicted dependence on $\theta_2$, but an unexpected and unexplained dependence on $\theta_1$. They elegantly showed a direct link between spin-waves and the fermion $s$CFS. Here we observe these predicted spin-waves in the resolved sideband regime, demonstrating the dependence of the $s$CFS on the correlations of the inhomogeneities of the two Ramsey pulses. The $s$CFS is almost invariably smaller for fermions as compared to bosons, generally non-zero, and vanishes for a pulse area $\theta_2$ near $\pi/2$, which maximizes the clock's stability.

The $s$CFS of a Fermi gas behaves distinctly differently than that of a Bose gas. The boson $s$CFS depends on the population difference of the two clocks states, often ascribed to the difference of mean-field energies. Thus, in Ramsey spectroscopy, the boson shift generally depends strongly on the area of the first pulse, which sets the populations for collisions that occur before the second Ramsey pulse. The boson $s$CFS is insensitive to the area of the second Ramsey pulse. For the fermion $s$CFS, the often-used mean field expression does not apply since the shift is in fact insensitive to $\theta_1$ ~\cite{Gibble2009}. It instead hinges on the atoms being excited differently so that they are distinguishable. Equally important is that the excitation inhomogeneity of the second Ramsey pulse is correlated with that of the first.  The atoms in each colliding pair experience opposite frequency shifts, which cancel on average if the second pulse is uncorrelated with the first.  Additionally, even for correlated pulses, if the second pulse reads out the phase shifts of both atoms with the same sensitivity (e.g. $\theta_2 \approx \pi/2$), the $s$CFS again vanishes~\cite{Gibble2009}.

These dependences on $\theta_1$ and $\theta_2$ were not observed in initial reports of Sr and Yb fermion $s$CFS's in optical lattice clocks~\cite{ Oates2009,Ludlow2009,Ye2009,Ye2011A,Rey2011A}.  The Yb shifts were actually $p$-wave and not $s$-wave~\cite{Ludlow2011}, apparent because they depended strongly on $\theta_1$ and not $\theta_2$, an explanation consistent with the Sr observations~\cite{Ludlow2009,Ye2009,Ye2011A,Rey2011A}.  Here, we additionally tune the $s$-wave scattering length through zero near a Feshbach resonance~\cite{Julienne05,*JuliennePrivate, Thomas2008} (Fig.~\ref{Fig1}(a)) to further and conclusively demonstrate that the shifts we detect are $s$-wave and not $p$-wave.

\begin{figure*}[hbtp]
\includegraphics[width=0.95\textwidth,angle=0,clip=true]{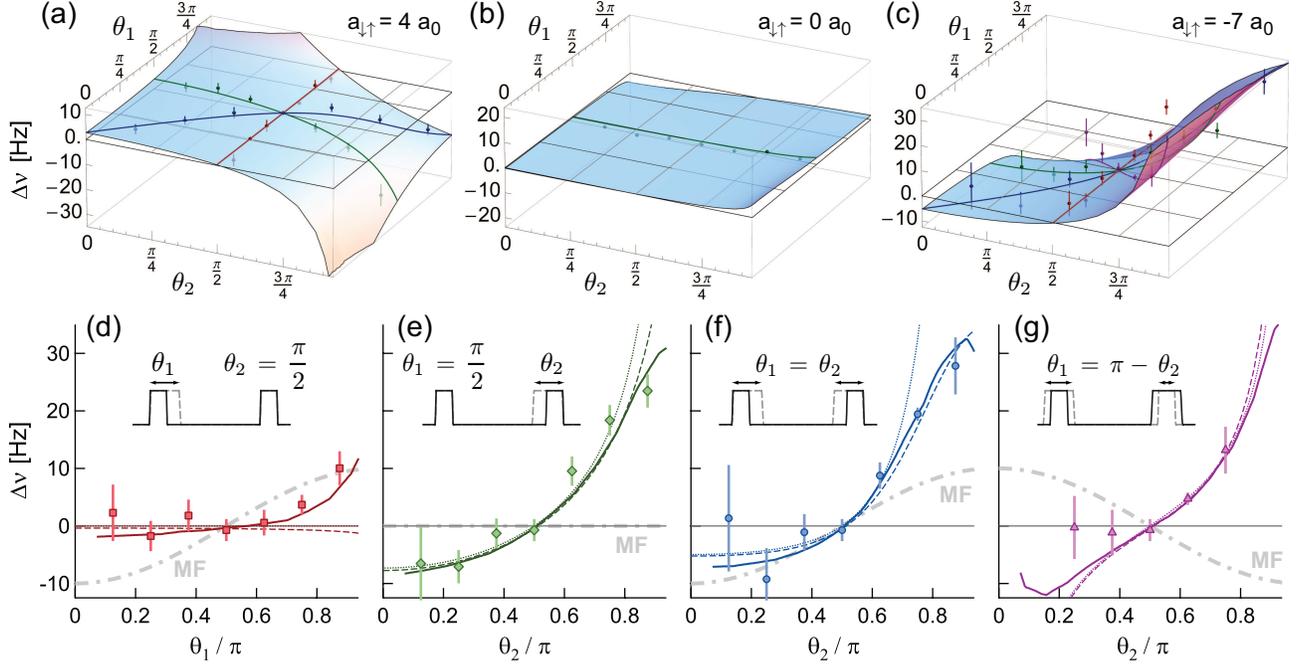}
\caption{(color online).  The $s$-wave collisional frequency shift of $^6$Li fermions versus Ramsey pulse areas ${\theta}_1$ and ${\theta}_2$.  (a) For $a_{\downarrow \uparrow} = 4 \, a_0$ ($B = 528.7\,{\mathrm{G}}$), the measured linear shift $\Delta\nu$ at $\rho_0= 8 \times 10^{12}\, {\mathrm{cm}^{-3}}$ is plotted along with Eq.~\ref{TwoAtomCFSWithSpinWvEq} for a thermal gas with $\Delta\Omega/\Omega = 0.2$, determined independently.  The green (red) solid line and points show the predicted and measured shift as $\theta_1$ ($\theta_2$) is varied with $\theta_2 (\theta_1) =\frac{\pi}{2}$.  The blue solid line and points show the dependence for equal pulse areas ${\theta}_1 = {\theta}_2$.  (b) and (c) are as in (a) with $a_{\downarrow \uparrow}  = 0 \, a_0$ ($B = 527.5 \, \mathrm{G}$) and $a_{\downarrow \uparrow} = - 7 \, a_0$ ($B = 525.5 \, {\mathrm{G}}$). The shift does vary strongly with $\theta_2$, going to 0 near $\theta_2 = \pi/2$.  (d -- g) The collisional frequency shift for $a_{\downarrow \uparrow} = - 7 \, a_0$ for the indicated surface cuts in (c).  Also shown are fits to the model for just two atoms (dotted), the model for many-atoms (dashed), and the numerically integrated model for many-atoms (solid), which includes the pulse shapes and Raman light shifts.  The shift is not proportional to the difference of partial densities which would imply a strong dependence on $\theta_1$ (grey dot-dashed mean-field curves).
}
\vspace{-0.01in}
\label{Fig2}
\end{figure*}

We use the two lowest-energy hyperfine states of $^6$Li atoms as clock states, denoted as $\left| \downarrow \right\rangle$ and $\left| \uparrow \right\rangle$.  At a bias field near 528~G the interstate $s$-wave scattering length $a_{\downarrow \uparrow} \simeq 0$ [Fig.~\ref{Fig1}(a)]~\cite{Julienne05,*JuliennePrivate,Thomas2008}.  The gas of $\sim 8 \times 10^4$ atoms has a temperature of $45\,\mu{\mathrm{K}}$, $\simeq 1.7 \times$ the Fermi temperature, and is confined in an optical dipole trap with trap frequencies $(\nu_x, \nu_y, \nu_z)  =  (3.8, 7.2, 11.0)\,{\mathrm{kHz}}$ [Fig.~\ref{Fig1}(b)].  We drive a two-photon Raman transition between $\left| \downarrow \right\rangle$ and $\left| \uparrow \right\rangle$ using copropagating laser fields, which are focused to $w_0 \simeq 10 \, \mu{\mathrm{m}}$, comparable to the thermal cloud radius $w_{\mathrm{th}} \simeq 6.5 \, \mu{\mathrm{m}}$ [Fig.~\ref{Fig1}(c)].  The Rabi frequency is inhomogeneous, giving different trap states different Rabi frequencies $\Omega_\alpha \propto \left\langle \psi_\alpha \right| e^{-2 r^2/w_0^2} \left| \psi_\alpha \right\rangle$.  The mean Rabi frequency for the ensemble, $\Omega \simeq 2 \pi \times 500\,{\mathrm{Hz}}$, is sufficiently below the trapping frequencies $\nu_y$ and  $\nu_z$.  A smooth turn-on and turn-off of the pulses further suppresses sideband excitations~\footnote{For each $\theta_1$, $\theta_2$, we adjust the interrogation time $T$ to keep a constant Ramsey fringe linewidth, with an effective interrogation time of $2.7 \, {\mathrm{ms}}$.  The Raman $\pi/2$ pulse duration is $450 \, \mu{\mathrm{s}}$. To always resolve the trap sidebands, we change the pulse area for $\theta > \pi/2$ ($\theta < \pi/2$) by increasing (decreasing) the duration (amplitude) of the Raman beam pulse and use a smooth turn-on and turn-off, which are $37.5 \, \mu{\mathrm{s}}$ long.   For $\theta < \pi/2$, the total optical intensity is maintained by adding an optical field off-resonant with the two-photon Raman transition.}.

To measure the $s$CFS, we record Ramsey fringes by shifting the phase of the second pulse with different pulse areas and densities.  Figure~\ref{Fig1}(d) shows the $s$CFS $\Delta \nu$ as a function of density for $\theta_1 = \frac{\pi}{2}$, $\theta_2 = \frac{3 \pi}{4}$, and several $a_{\downarrow \uparrow}$.  The apparent pulse areas ${\theta}_1$ and ${\theta}_2$ are defined as ${\theta}_1 = {\theta}_2 = \pi/2$ giving maximum Ramsey fringe contrast.  The mean total density, $\rho = \rho_\downarrow + \rho_\uparrow $, is normalized to our canonical maximum of the mean density, $\rho_0 = 8 \times 10^{12}\, {\mathrm{cm}^{-3}}$.  With a large $a_{\downarrow \uparrow}$, the density dependence of $\Delta \nu$ is nonlinear at high density (see below).  To extract the $s$CFS for a weakly interacting Fermi gas, we vary $\theta_1$ and $\theta_2$ while at low density and small $a_{\downarrow \uparrow}$ [Fig.~\ref{Fig1}(d) inset], where $\Delta \nu$ is linear in $\rho$ and $a_{\downarrow \uparrow}$.

The predicted $s$CFS of a weakly-interacting, many-body Fermi gas emerges when just two atoms are considered~\cite{Gibble2009}. The singlet-triplet basis for the two-atom wavefunction [Fig.~\ref{Fig1}(e)] is helpful because only the singlet state has a collisional interaction $g$, proportional to $a_{\downarrow \uparrow}$~\footnote{We consider a contact interaction $V({\bf{r}}_1,{\bf{r}}_2) = \frac{4 \pi \hbar^2 a_{\downarrow \uparrow}}{m} \, \delta({\bf{r}}_1 - {\bf{r}}_2)$. Here, $g = \frac{2 h a_{\downarrow \uparrow}}{m} \int \psi_\alpha ^2({\bf{r}}) \, \psi_\beta^2({\bf{r}}) \, dV$.}.  The triplet (singlet) states are the product of a symmetric (antisymmetric) spin wavefunction and an antisymmetric (symmetric) combination of spatial wavefunctions, trap-eigenstates $\psi_\alpha $ and $\psi_\beta$ [Fig.~\ref{Fig1}(e)].  We initially prepare the atoms in the same internal state, $\downarrow$, and illuminate them with two spatially inhomogeneous Ramsey pulses [Fig.~\ref{Fig1}(c)], giving different Rabi frequencies, $\Omega_\alpha $ and $\Omega_\beta$.  Because the atoms in the different trap states have different Rabi frequencies, they become distinguishable, populate a pair-wise singlet state, and interact during the interrogation time $T$ between the two Ramsey pulses~\footnote{We use scattering lengths much smaller than the deBroglie wavelength, $2000 a_0$, so that trap state changing (lateral) collisions can be neglected.}.  For weak interactions, $g T \ll 1$, the $s$CFS for two or more atoms is~\cite{Gibble2009}:
\begin{eqnarray}
\Delta \nu & = & \frac{{\displaystyle{\sum\limits_{{\mathrm{pairs}}}{ g \, T \,\sin ( 2 \Delta \theta_1) \, \sin ( \Delta \theta_2 ) \, \cos(\bar{\theta}_2)}}}}{ \pi T \mathcal{A}}, \quad \label{TwoAtomCFSWithSpinWvEq}
\end{eqnarray}
where the sum is over all atom pairs.  Here, $\Delta \theta_{i} = \Delta \Omega \, \tau_{i}$ and $\bar{\theta}_{i} = \bar{\Omega} \, \tau_{i}$,  where $\tau_{1 \, (2)}$ denotes the first (second) pulse duration, $\bar{\Omega} \equiv \left( \Omega_\alpha + \Omega_\beta \right)/2 \gg g$, and $\Delta\Omega \equiv \left( \Omega_\alpha - \Omega_\beta \right)/2$. The Ramsey fringe amplitude $\mathcal{A}$ is $\sum_k \sin(\theta_{1,k}) \sin(\theta_{2,k})$ for weak interactions.

\begin{figure}[htbp]
\includegraphics[width=\columnwidth,angle=0,clip=true]{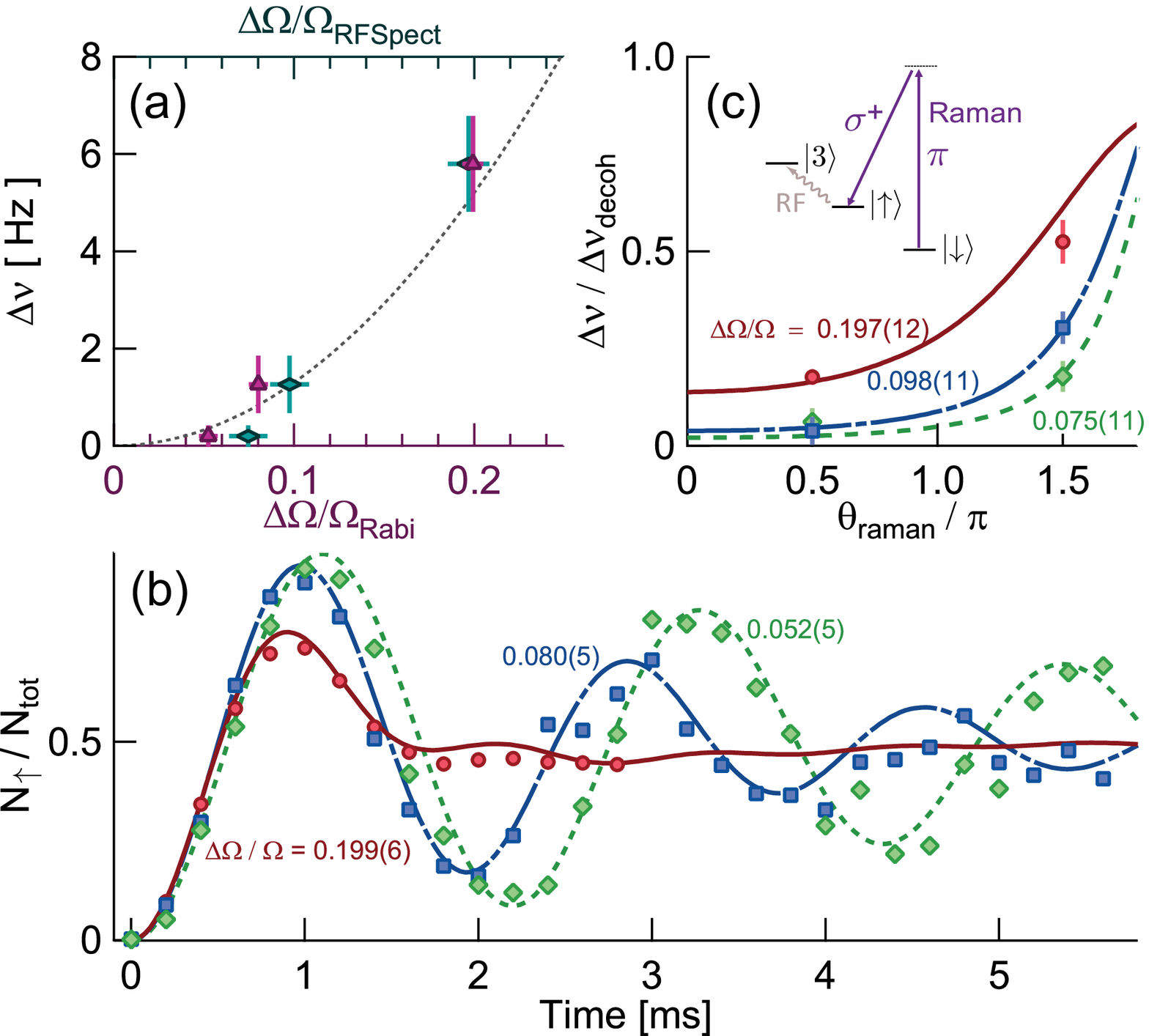}
\caption{(color online). (a) The measured $s$CFS grows with increasing inhomogeneity $\Delta\Omega$, independently measured with Rabi oscillations (${\color{MyPurple}\blacktriangle}$) as in (b) and the frequency shift of the $\left| \uparrow \right\rangle - \left| 3 \right\rangle$ RF transition(\begin{sideways}${\color{MyTurquoise}\blacklozenge}$\end{sideways}) as in (c).  The scattering length $a_{\downarrow \uparrow}$ is $-7 \, a_0$ and the dashed curve shows the expected quadratic dependence.  (b) Decay of Rabi oscillations for Raman beams with waists comparable to the radius of the thermal cloud.  A fit of the Rabi oscillations of a trapped thermal gas determines $\Delta\Omega$.  (c) We use the collisional frequency shift of the $\left| \uparrow \right\rangle - \left| 3 \right\rangle$ transition, with a spatially homogeneous RF Rabi pulse, to measure the $\left| \downarrow \right\rangle$ -- $\left| \uparrow \right\rangle$ singlet state population. The shift increases as expected with the inhomogeneity of the $\left| \downarrow \right\rangle - \left| \uparrow \right\rangle$ Raman field and, via a model, also yields $\Delta\Omega$. }
\label{Fig3}
\end{figure}

In Fig.~\ref{Fig2}(a) we plot the prediction from Eq.~\ref{TwoAtomCFSWithSpinWvEq}  and the observed linear shift $\Delta \nu$ at $\rho_0 = 8 \times 10^{12}\, {\mathrm{cm}^{-3}}$, $a_{\downarrow \uparrow} = 4 \, a_0$, and $\Delta\Omega/ \Omega = 0.20$~\footnote{We define $\Delta \Omega^2 \equiv  {\langle} g_{\alpha \beta} \Delta\Omega_{\alpha \beta}^2 {\rangle} /{\langle} g_{\alpha \beta} {\rangle}$, where ${\langle} \, {\rangle}$ denotes an ensemble average over all atom pairs in states $\psi_\alpha$ and $\psi_\beta$. Then, for small $\Delta \Omega$ and $\Omega$, $\Delta \nu$ is $ (g/\pi) \, (\Delta \Omega/\Omega)^2$ for two atoms and $( (N - 1) {\langle} g_{\alpha \beta}{\rangle}/\pi) \, ( \Delta \Omega/\Omega)^2$ for $N$ atoms. Here, $\Omega$ is the average Rabi frequency.  In the context of the Rabi flopping in Fig.~\ref{Fig3}(b), a reasonable definition of $\Delta\Omega^2$ is simply ${\langle} \Delta\Omega_{\alpha \beta}^2 {\rangle}$, but this neglects important correlations between $\Delta\Omega_{\alpha \beta}^2$ and ${g}_{\alpha \beta}$. }\MakeFootnoteRef{Note_DOmega}.  Two independent methods determine $\Delta \Omega$ and are described below.  The $\Delta \nu$ surface in Fig.~\ref{Fig2}(a) is Eq.~\ref{TwoAtomCFSWithSpinWvEq} for a thermal gas with a single-parameter fit of the amplitude, to account for our uncertainty in the absolute atom number $N$ ($\simeq 25 \%$).  Versus $B$, $\Delta\nu$  gives $a_{\downarrow \uparrow} = \left[ 3.5(9)\,a_0/{\mathrm{G}} \right] \,  \left[B - 527.61(7)\,{\mathrm{G}} \right]$ (Fig.~\ref{Fig1}(a)), in agreement with Refs.~\cite{Julienne05,*JuliennePrivate,Thomas2008}.

The $s$CFS given by Eq.~\ref{TwoAtomCFSWithSpinWvEq} vanishes if either $\Delta\theta_1$ or $\Delta\theta_2$ is zero.  Physically, $\Delta \theta_1$ must be nonzero to make the atoms distinguishable and populate the singlet state. During the interrogation time $T$, the singlet state acquires a collisional phase shift $\exp[2i g T]$ and the second pulse must also be inhomogeneous so that this phase shift is read out. For $\bar{\theta}_2 = \pi/2$ (Fig.~\ref{Fig2}(a -- c \& e), green curves and data), $\cos(\bar{\theta}_2)$ goes to 0 so that the $s$CFS vanishes, even for inhomogeneous clock fields, because the opposite phase shifts of each atom are read out with equal sensitivities~\cite{Gibble2009}.

It is remarkable that $\Delta \nu$ in Eq.~\ref{TwoAtomCFSWithSpinWvEq}  is insensitive to $\bar{\theta}_1$ (Fig.~\ref{Fig2}(a, c, \& d), red curves and data).  The amplitude in the singlet state (which acquires the collisional phase shift) depends only on $\Delta\theta_1$ and not on $\bar{\theta}_1$.  The $s$CFS is in general non-zero for $\bar{\theta}_1 = \pi/2$, for which the population difference $N_\uparrow - N_\downarrow = - 2 N \cos(\bar{\theta}_1) = 0$ during the interrogation time $T$.  This is in stark contrast with the widely-used mean-field expression (grey dot-dashed curves in Fig.~\ref{Fig2} (d - g))~\cite{Cornell2002,Ketterle2003A,Ketterle2003B,Ludlow2009,Oates2009,Ye2009,Ye2011A},  where the $s$CFS $\Delta \nu$ vanishes at $N_\uparrow - N_\downarrow = 0$, and disagrees with an interpretation of Rabi spectroscopy of a Fermi gas~\cite{Rey2009}.  A key result of this work is the experimental measurement of the fermion $s$CFS, including an insensitivity of the $s$CFS to $\bar{\theta}_1$.

Tuning $a_{\downarrow \uparrow}$ to $0$ in Fig.~\ref{Fig2}(b) gives no $s$CFS as expected.  With $a_{\downarrow \uparrow} = - 7 \,a_0$ , the $s$CFS in Fig.~\ref{Fig2}(c) has the opposite sign of Fig.~\ref{Fig2}(a).  The fit to Eq.~\ref{TwoAtomCFSWithSpinWvEq} (dashed line) for an ensemble with $a_{\downarrow \uparrow} = -7 \, a_0$ is also shown, along with the model for just two atoms (dotted)  with an effective $g$ chosen to give the same $\Delta \nu$ for $\theta_{1,2}\rightarrow 0$.  Equation~\ref{TwoAtomCFSWithSpinWvEq} uses a short-pulse approximation where only interactions during the interrogation time are considered.  In the experiment, the atoms also acquire a phase shift due to interactions during the pulses.  Further, the Raman beams produce a small time-dependent light shift~\footnote{The Raman beams are detuned to the red (blue) of the D2 (D1) line by 4.8 GHz (5.2 GHz) to minimize spontaneous emission and the differential light shift $\Delta\nu_{\protect \mathrm{LS}}$.  For a trap state $\psi_\alpha$,  $\Delta\nu_{ \protect \mathrm{LS}}$ is $ \simeq 0.34 \, \Omega_\alpha/(2 \pi)$.  Weak laser beams clear the small fraction of atoms in states populated by spontaneous emission to eliminate extraneous collisional shifts.} of the transition during each pulse.  We include these effects and the pulse shapes in a numerically integrated Monte-Carlo simulation (solid lines in Fig.~\ref{Fig2}(d - g)).  All of these give small corrections, and the interactions during the pulses are the largest of these.

We highlight that correlations between $g_{\alpha \beta}$, and $\theta_{\alpha, \beta}$ shift the pulse area where $\Delta\nu = 0$ from $\theta_2 = \pi/2$ to $\theta_2 = 0.51 \, \pi$ in Fig.~\ref{Fig2}(e).  Here, the average of $\cos\bar{\theta}_2$ crosses zero at $\theta_2 = 0.56\,\pi$ because the Ramsey fringe contrast has bigger contributions from large $\theta_{\alpha}$.  However, atoms in low vibrational states have large $\bar{\theta}_2$ and also large $g_{\alpha \beta}$ since they reside in the high density region of the trap. This pushes the zero-crossing of $\left\langle g_{\alpha \beta} \, \cos\bar{\theta}_2 \right\rangle$ to smaller $\theta_2$.  Finally, because trap states that have a large overlap with each other have similar overlaps with the clock field, $\Delta\theta_1 \, \Delta\theta_2$ and $g_{\alpha \beta}$ are anti-correlated and this pushes the zero-crossing to larger $\theta_2$.  These perturbations are quite general, for both amplitude variations and wavefront tilts, and in detail depend on the excitation inhomogeneity and trap geometry.

With a large $a_{\downarrow \uparrow}$, the density dependence of $\Delta \nu$ is nonlinear at high density (Fig.~\ref{Fig1}(d)) and reverses sign once the acquired phase shift during the interrogation time $T$ exceeds $\pi$.  For just two atoms, Eq.~\ref{TwoAtomCFSWithSpinWvEq} has a sinusoidal dependence $\sin (2 g T)$, instead of a linear dependence on $g T$,  and this gives the observed sinusoidal dependence on density, where the effective $g$ for many atoms is proportional to density~\cite{Gibble2009, Rosenbusch2012}.  As in Ref. \citenum{Rosenbusch2012}, but here in the resolved-sideband limit, the sinusoidal dependence is a manifestation of a spin wave, although we do not spatially resolve the oscillating spin populations.

Fig.~\ref{Fig3}(a) shows that the $s$CFS increases quadratically with the spatial inhomogeneity. Here, we vary the Raman beam waist, measuring the $s$CFS versus $\theta_{1,2}$ with $a_{\downarrow \uparrow}  = -7 a_0$. The fitted surface, in the limit of small $\theta_{1,2}$,  gives $\Delta\nu$, which is plotted versus $\Delta\Omega/\Omega$.  We experimentally determine $\Delta\Omega$ in two independent ways. First, we drive the Raman transition and fit the decaying Rabi oscillations between $\left| \downarrow \right\rangle$ and $\left| \uparrow \right\rangle$ (Fig.~\ref{Fig3}(b)).  The circles, squares, and diamonds show the Rabi oscillations for increasing beam waists.  We model the Rabi oscillations of a thermal distribution of trap states $\psi_\alpha$ with different Rabi frequencies $\Omega_\alpha$.  Second, we probe the interactions with RF spectroscopy to a third internal state $\left| 3 \right\rangle$~\cite{Ketterle2003A,Ketterle2003B,Jin2003B}.  A Raman pulse of area $\theta_1$ is first applied to a sample prepared in $\left| \downarrow \right\rangle$ to populate pairwise singlet states, producing an interaction shift.  This shift is observed with Rabi spectroscopy on the $\left| \uparrow \right\rangle - \left| 3 \right\rangle$ transition.  The interaction shift $\Delta \nu_{\uparrow 3}$ is proportional to $\sum_{\alpha \neq \beta} \left(g_{\alpha \beta, \downarrow \uparrow} - g_{ \alpha \beta,\uparrow 3}\right) \sin^2 \Delta\theta_{1, \alpha \beta}$.  It increases with the singlet state population, $\sin^2 \Delta\theta_{1, \alpha \beta}$, which increases as $\Delta\Omega$ increases.  The circles, squares and diamonds in Fig.~\ref{Fig3}(c) shows the measured shift versus $\theta_1$, normalized to the shift of a fully decohered gas.  The fully decohered gas is prepared by optical pumping to create an equal $\downarrow \, \uparrow$ mixture.  The curves are single-parameter fits of $\Delta\Omega/\Omega$.  The dashed curve in Fig.~\ref{Fig3}(a) shows the expected quadratic dependence of $\Delta\nu$ on $\Delta\Omega$.

Our experiment elucidates the novel behaviors of the collisional frequency shifts of clocks based on fermions.  For both Ramsey and Rabi spectroscopy, the spatial inhomogeneities of the clock field make fermions distinguishable, producing a shift if the inhomogeneities of the excitation and the readout of the phases of the clock coherences are correlated. Optical lattice clocks are naturally susceptible to these shifts since optical-frequency clock fields have fast spatial variations.  Wavefront curvatures or small tilts of the clock-field phase fronts lead to different Rabi frequencies in a gas. When the $p$-wave collisions are not frozen out and the scattering lengths are favorable, both the $p$- and $s$-wave collisional frequency shifts can be independently canceled by adjusting $\theta_1$ and $\theta_2$ around $\pi/2$.  We note that correlations in the sample generally perturb the zero-crossing of the $s$-wave collisional frequency shift to a $\theta_2$ slightly different from $\pi/2$.  The shifts for fermions are naturally much smaller than those for clocks based on bosons, demonstrating that ultracold fermions are excellent candidates for a variety of precision experiments.


\begin{acknowledgments}
The authors acknowledge helpful discussions with W. Maineult and P. Rosenbusch and support from the NSF (Grant Nos.~PHY-1011156, PHY-0800233, PHY-1209662), the AFOSR (Grant No.~FA9550-08-1-0069), the ARO (Grant No.~W911NF-06-1-0398, which included partial funding from the DARPA OLE program), the Sloan Foundation (ELH), and Penn State.
\end{acknowledgments}


%


\begin{thebibliography}{32}%
\makeatletter
\providecommand \@ifxundefined [1]{%
 \@ifx{#1\undefined}
}%
\providecommand \@ifnum [1]{%
 \ifnum #1\expandafter \@firstoftwo
 \else \expandafter \@secondoftwo
 \fi
}%
\providecommand \@ifx [1]{%
 \ifx #1\expandafter \@firstoftwo
 \else \expandafter \@secondoftwo
 \fi
}%
\providecommand \natexlab [1]{#1}%
\providecommand \enquote  [1]{``#1''}%
\providecommand \bibnamefont  [1]{#1}%
\providecommand \bibfnamefont [1]{#1}%
\providecommand \citenamefont [1]{#1}%
\providecommand \href@noop [0]{\@secondoftwo}%
\providecommand \href [0]{\begingroup \@sanitize@url \@href}%
\providecommand \@href[1]{\@@startlink{#1}\@@href}%
\providecommand \@@href[1]{\endgroup#1\@@endlink}%
\providecommand \@sanitize@url [0]{\catcode `\\12\catcode `\$12\catcode
  `\&12\catcode `\#12\catcode `\^12\catcode `\_12\catcode `\%12\relax}%
\providecommand \@@startlink[1]{}%
\providecommand \@@endlink[0]{}%
\providecommand \url  [0]{\begingroup\@sanitize@url \@url }%
\providecommand \@url [1]{\endgroup\@href {#1}{\urlprefix }}%
\providecommand \urlprefix  [0]{URL }%
\providecommand \Eprint [0]{\href }%
\providecommand \doibase [0]{http://dx.doi.org/}%
\providecommand \selectlanguage [0]{\@gobble}%
\providecommand \bibinfo  [0]{\@secondoftwo}%
\providecommand \bibfield  [0]{\@secondoftwo}%
\providecommand \translation [1]{[#1]}%
\providecommand \BibitemOpen [0]{}%
\providecommand \bibitemStop [0]{}%
\providecommand \bibitemNoStop [0]{.\EOS\space}%
\providecommand \EOS [0]{\spacefactor3000\relax}%
\providecommand \BibitemShut  [1]{\csname bibitem#1\endcsname}%
\let\auto@bib@innerbib\@empty
\bibitem [{\citenamefont {Guena}\ \emph {et~al.}(2012)\citenamefont {Guena},
  \citenamefont {Abgrall}, \citenamefont {Rovera}, \citenamefont {Laurent},
  \citenamefont {Chupin}, \citenamefont {Lours}, \citenamefont {Santarelli},
  \citenamefont {Rosenbusch}, \citenamefont {Tobar}, \citenamefont {Li},
  \citenamefont {Gibble}, \citenamefont {Clairon},\ and\ \citenamefont
  {Bize}}]{Bize2012}%
  \BibitemOpen
  \bibfield  {author} {\bibinfo {author} {\bibfnamefont {J.}~\bibnamefont
  {Guena}}, \bibinfo {author} {\bibfnamefont {M.}~\bibnamefont {Abgrall}},
  \bibinfo {author} {\bibfnamefont {D.}~\bibnamefont {Rovera}}, \bibinfo
  {author} {\bibfnamefont {P.}~\bibnamefont {Laurent}}, \bibinfo {author}
  {\bibfnamefont {B.}~\bibnamefont {Chupin}}, \bibinfo {author} {\bibfnamefont
  {M.}~\bibnamefont {Lours}}, \bibinfo {author} {\bibfnamefont
  {G.}~\bibnamefont {Santarelli}}, \bibinfo {author} {\bibfnamefont
  {P.}~\bibnamefont {Rosenbusch}}, \bibinfo {author} {\bibfnamefont {M.~E.}\
  \bibnamefont {Tobar}}, \bibinfo {author} {\bibfnamefont {R.}~\bibnamefont
  {Li}}, \bibinfo {author} {\bibfnamefont {K.}~\bibnamefont {Gibble}}, \bibinfo
  {author} {\bibfnamefont {A.}~\bibnamefont {Clairon}}, \ and\ \bibinfo
  {author} {\bibfnamefont {S.}~\bibnamefont {Bize}},\ }\href@noop {} {\bibfield
   {journal} {\bibinfo  {journal} {IEEE Trans. Ultrason. Ferroelectr. Freq.
  Control}\ }\textbf {\bibinfo {volume} {59}},\ \bibinfo {pages} {391}
  (\bibinfo {year} {2012})}\BibitemShut {NoStop}%
\bibitem [{\citenamefont {Li}\ \emph {et~al.}(2011)\citenamefont {Li},
  \citenamefont {Gibble},\ and\ \citenamefont {Szymaniec}}]{Szymaniec2011}%
  \BibitemOpen
  \bibfield  {author} {\bibinfo {author} {\bibfnamefont {R.}~\bibnamefont
  {Li}}, \bibinfo {author} {\bibfnamefont {K.}~\bibnamefont {Gibble}}, \ and\
  \bibinfo {author} {\bibfnamefont {K.}~\bibnamefont {Szymaniec}},\ }\href@noop
  {} {\bibfield  {journal} {\bibinfo  {journal} {Metrologia}\ }\textbf
  {\bibinfo {volume} {48}},\ \bibinfo {pages} {283} (\bibinfo {year}
  {2011})}\BibitemShut {NoStop}%
\bibitem [{\citenamefont {Zwierlein}\ \emph {et~al.}(2003)\citenamefont
  {Zwierlein}, \citenamefont {Hadzibabic}, \citenamefont {Gupta},\ and\
  \citenamefont {Ketterle}}]{Ketterle2003B}%
  \BibitemOpen
  \bibfield  {author} {\bibinfo {author} {\bibfnamefont {M.~W.}\ \bibnamefont
  {Zwierlein}}, \bibinfo {author} {\bibfnamefont {Z.}~\bibnamefont
  {Hadzibabic}}, \bibinfo {author} {\bibfnamefont {S.}~\bibnamefont {Gupta}}, \
  and\ \bibinfo {author} {\bibfnamefont {W.}~\bibnamefont {Ketterle}},\
  }\href@noop {} {\bibfield  {journal} {\bibinfo  {journal} {Phys. Rev. Lett.}\
  }\textbf {\bibinfo {volume} {91}},\ \bibinfo {pages} {250404} (\bibinfo
  {year} {2003})}\BibitemShut {NoStop}%
\bibitem [{\citenamefont {Gibble}\ and\ \citenamefont
  {Verhaar}(1995)}]{Verhaar1995}%
  \BibitemOpen
  \bibfield  {author} {\bibinfo {author} {\bibfnamefont {K.}~\bibnamefont
  {Gibble}}\ and\ \bibinfo {author} {\bibfnamefont {B.~J.}\ \bibnamefont
  {Verhaar}},\ }\href@noop {} {\bibfield  {journal} {\bibinfo  {journal} {Phys.
  Rev. A}\ }\textbf {\bibinfo {volume} {52}},\ \bibinfo {pages} {3370}
  (\bibinfo {year} {1995})}\BibitemShut {NoStop}%
\bibitem [{\citenamefont {Gupta}\ \emph {et~al.}(2003)\citenamefont {Gupta},
  \citenamefont {Hadzibabic}, \citenamefont {Zwierlein}, \citenamefont {Stan},
  \citenamefont {Dieckmann}, \citenamefont {Schunck}, \citenamefont {van
  Kempen}, \citenamefont {Verhaar},\ and\ \citenamefont
  {Ketterle}}]{Ketterle2003A}%
  \BibitemOpen
  \bibfield  {author} {\bibinfo {author} {\bibfnamefont {S.}~\bibnamefont
  {Gupta}}, \bibinfo {author} {\bibfnamefont {Z.}~\bibnamefont {Hadzibabic}},
  \bibinfo {author} {\bibfnamefont {M.}~\bibnamefont {Zwierlein}}, \bibinfo
  {author} {\bibfnamefont {C.}~\bibnamefont {Stan}}, \bibinfo {author}
  {\bibfnamefont {K.}~\bibnamefont {Dieckmann}}, \bibinfo {author}
  {\bibfnamefont {C.}~\bibnamefont {Schunck}}, \bibinfo {author} {\bibfnamefont
  {E.}~\bibnamefont {van Kempen}}, \bibinfo {author} {\bibfnamefont
  {B.}~\bibnamefont {Verhaar}}, \ and\ \bibinfo {author} {\bibfnamefont
  {W.}~\bibnamefont {Ketterle}},\ }\href@noop {} {\bibfield  {journal}
  {\bibinfo  {journal} {Science}\ }\textbf {\bibinfo {volume} {300}},\ \bibinfo
  {pages} {1723} (\bibinfo {year} {2003})}\BibitemShut {NoStop}%
\bibitem [{\citenamefont {Takamoto}\ \emph {et~al.}(2005)\citenamefont
  {Takamoto}, \citenamefont {Hong}, \citenamefont {Higashi},\ and\
  \citenamefont {Katori}}]{Katori2005}%
  \BibitemOpen
  \bibfield  {author} {\bibinfo {author} {\bibfnamefont {M.}~\bibnamefont
  {Takamoto}}, \bibinfo {author} {\bibfnamefont {F.-L.}\ \bibnamefont {Hong}},
  \bibinfo {author} {\bibfnamefont {R.}~\bibnamefont {Higashi}}, \ and\
  \bibinfo {author} {\bibfnamefont {H.}~\bibnamefont {Katori}},\ }\href@noop {}
  {\bibfield  {journal} {\bibinfo  {journal} {Nature}\ }\textbf {\bibinfo
  {volume} {435}},\ \bibinfo {pages} {321} (\bibinfo {year}
  {2005})}\BibitemShut {NoStop}%
\bibitem [{\citenamefont {Roati}\ \emph {et~al.}(2004)\citenamefont {Roati},
  \citenamefont {de~Mirandes}, \citenamefont {Ferlaino}, \citenamefont {Ott},
  \citenamefont {Modugno},\ and\ \citenamefont {Inguscio}}]{Inguscio2004}%
  \BibitemOpen
  \bibfield  {author} {\bibinfo {author} {\bibfnamefont {G.}~\bibnamefont
  {Roati}}, \bibinfo {author} {\bibfnamefont {E.}~\bibnamefont {de~Mirandes}},
  \bibinfo {author} {\bibfnamefont {F.}~\bibnamefont {Ferlaino}}, \bibinfo
  {author} {\bibfnamefont {H.}~\bibnamefont {Ott}}, \bibinfo {author}
  {\bibfnamefont {G.}~\bibnamefont {Modugno}}, \ and\ \bibinfo {author}
  {\bibfnamefont {M.}~\bibnamefont {Inguscio}},\ }\href {\doibase
  10.1103/PhysRevLett.92.230402} {\bibfield  {journal} {\bibinfo  {journal}
  {Phys. Rev. Lett.}\ }\textbf {\bibinfo {volume} {92}},\ \bibinfo {pages}
  {230402} (\bibinfo {year} {2004})}\BibitemShut {NoStop}%
\bibitem [{\citenamefont {Budker}\ and\ \citenamefont
  {Romalis}(2007)}]{Romalis2007}%
  \BibitemOpen
  \bibfield  {author} {\bibinfo {author} {\bibfnamefont {D.}~\bibnamefont
  {Budker}}\ and\ \bibinfo {author} {\bibfnamefont {M.}~\bibnamefont
  {Romalis}},\ }\href {\doibase 10.1038/nphys566} {\bibfield  {journal}
  {\bibinfo  {journal} {Nature Physics}\ }\textbf {\bibinfo {volume} {3}},\
  \bibinfo {pages} {227} (\bibinfo {year} {2007})}\BibitemShut {NoStop}%
\bibitem [{\citenamefont {Liu}\ \emph {et~al.}(2001)\citenamefont {Liu},
  \citenamefont {Dutton}, \citenamefont {Behroozi},\ and\ \citenamefont
  {Hau}}]{Hau2001}%
  \BibitemOpen
  \bibfield  {author} {\bibinfo {author} {\bibfnamefont {C.}~\bibnamefont
  {Liu}}, \bibinfo {author} {\bibfnamefont {Z.}~\bibnamefont {Dutton}},
  \bibinfo {author} {\bibfnamefont {C.}~\bibnamefont {Behroozi}}, \ and\
  \bibinfo {author} {\bibfnamefont {L.}~\bibnamefont {Hau}},\ }\href@noop {}
  {\bibfield  {journal} {\bibinfo  {journal} {Nature}\ }\textbf {\bibinfo
  {volume} {409}},\ \bibinfo {pages} {490} (\bibinfo {year}
  {2001})}\BibitemShut {NoStop}%
\bibitem [{\citenamefont {Chou}\ \emph {et~al.}(2005)\citenamefont {Chou},
  \citenamefont {de~Riedmatten}, \citenamefont {Felinto}, \citenamefont
  {Polyakov}, \citenamefont {van Enk},\ and\ \citenamefont
  {Kimble}}]{Kimble2005}%
  \BibitemOpen
  \bibfield  {author} {\bibinfo {author} {\bibfnamefont {C.}~\bibnamefont
  {Chou}}, \bibinfo {author} {\bibfnamefont {H.}~\bibnamefont {de~Riedmatten}},
  \bibinfo {author} {\bibfnamefont {D.}~\bibnamefont {Felinto}}, \bibinfo
  {author} {\bibfnamefont {S.}~\bibnamefont {Polyakov}}, \bibinfo {author}
  {\bibfnamefont {S.}~\bibnamefont {van Enk}}, \ and\ \bibinfo {author}
  {\bibfnamefont {H.}~\bibnamefont {Kimble}},\ }\href@noop {} {\bibfield
  {journal} {\bibinfo  {journal} {Nature}\ }\textbf {\bibinfo {volume} {438}},\
  \bibinfo {pages} {828} (\bibinfo {year} {2005})}\BibitemShut {NoStop}%
\bibitem [{\citenamefont {Chaneliere}\ \emph {et~al.}(2005)\citenamefont
  {Chaneliere}, \citenamefont {Matsukevich}, \citenamefont {Jenkins},
  \citenamefont {Lan}, \citenamefont {Kennedy},\ and\ \citenamefont
  {Kuzmich}}]{Kuzmich2005}%
  \BibitemOpen
  \bibfield  {author} {\bibinfo {author} {\bibfnamefont {T.}~\bibnamefont
  {Chaneliere}}, \bibinfo {author} {\bibfnamefont {D.}~\bibnamefont
  {Matsukevich}}, \bibinfo {author} {\bibfnamefont {S.}~\bibnamefont
  {Jenkins}}, \bibinfo {author} {\bibfnamefont {S.}~\bibnamefont {Lan}},
  \bibinfo {author} {\bibfnamefont {T.}~\bibnamefont {Kennedy}}, \ and\
  \bibinfo {author} {\bibfnamefont {A.}~\bibnamefont {Kuzmich}},\ }\href@noop
  {} {\bibfield  {journal} {\bibinfo  {journal} {Nature}\ }\textbf {\bibinfo
  {volume} {438}},\ \bibinfo {pages} {833} (\bibinfo {year}
  {2005})}\BibitemShut {NoStop}%
\bibitem [{\citenamefont {Bao}\ \emph {et~al.}(2012)\citenamefont {Bao},
  \citenamefont {Reingruber}, \citenamefont {Dietrich}, \citenamefont {Rui},
  \citenamefont {Dueck}, \citenamefont {Strassel}, \citenamefont {Li},
  \citenamefont {Liu}, \citenamefont {Zhao},\ and\ \citenamefont
  {Pan}}]{JianWei2012}%
  \BibitemOpen
  \bibfield  {author} {\bibinfo {author} {\bibfnamefont {X.-H.}\ \bibnamefont
  {Bao}}, \bibinfo {author} {\bibfnamefont {A.}~\bibnamefont {Reingruber}},
  \bibinfo {author} {\bibfnamefont {P.}~\bibnamefont {Dietrich}}, \bibinfo
  {author} {\bibfnamefont {J.}~\bibnamefont {Rui}}, \bibinfo {author}
  {\bibfnamefont {A.}~\bibnamefont {Dueck}}, \bibinfo {author} {\bibfnamefont
  {T.}~\bibnamefont {Strassel}}, \bibinfo {author} {\bibfnamefont
  {L.}~\bibnamefont {Li}}, \bibinfo {author} {\bibfnamefont {N.-L.}\
  \bibnamefont {Liu}}, \bibinfo {author} {\bibfnamefont {B.}~\bibnamefont
  {Zhao}}, \ and\ \bibinfo {author} {\bibfnamefont {J.-W.}\ \bibnamefont
  {Pan}},\ }\href@noop {} {\bibfield  {journal} {\bibinfo  {journal} {Nature
  Physics}\ }\textbf {\bibinfo {volume} {8}},\ \bibinfo {pages} {517} (\bibinfo
  {year} {2012})}\BibitemShut {NoStop}%
\bibitem [{\citenamefont {Gibble}(2009)}]{Gibble2009}%
  \BibitemOpen
  \bibfield  {author} {\bibinfo {author} {\bibfnamefont {K.}~\bibnamefont
  {Gibble}},\ }\href@noop {} {\bibfield  {journal} {\bibinfo  {journal} {Phys.
  Rev. Lett.}\ }\textbf {\bibinfo {volume} {103}},\ \bibinfo {pages} {113202}
  (\bibinfo {year} {2009})}\BibitemShut {NoStop}%
\bibitem [{\citenamefont {Rey}\ \emph {et~al.}(2009)\citenamefont {Rey},
  \citenamefont {Gorshkov},\ and\ \citenamefont {Rubbo}}]{Rey2009}%
  \BibitemOpen
  \bibfield  {author} {\bibinfo {author} {\bibfnamefont {A.~M.}\ \bibnamefont
  {Rey}}, \bibinfo {author} {\bibfnamefont {A.~V.}\ \bibnamefont {Gorshkov}}, \
  and\ \bibinfo {author} {\bibfnamefont {C.}~\bibnamefont {Rubbo}},\
  }\href@noop {} {\bibfield  {journal} {\bibinfo  {journal} {Phys. Rev. Lett.}\
  }\textbf {\bibinfo {volume} {103}},\ \bibinfo {pages} {260402} (\bibinfo
  {year} {2009})}\BibitemShut {NoStop}%
\bibitem [{\citenamefont {Yu}\ and\ \citenamefont
  {Pethick}(2010)}]{Pethick2010}%
  \BibitemOpen
  \bibfield  {author} {\bibinfo {author} {\bibfnamefont {Z.}~\bibnamefont
  {Yu}}\ and\ \bibinfo {author} {\bibfnamefont {C.~J.}\ \bibnamefont
  {Pethick}},\ }\href@noop {} {\bibfield  {journal} {\bibinfo  {journal} {Phys.
  Rev. Lett.}\ }\textbf {\bibinfo {volume} {104}},\ \bibinfo {pages} {010801}
  (\bibinfo {year} {2010})}\BibitemShut {NoStop}%
\bibitem [{\citenamefont {Maineult}\ \emph {et~al.}(2012)\citenamefont
  {Maineult}, \citenamefont {Deutsch}, \citenamefont {Gibble}, \citenamefont
  {Reichel},\ and\ \citenamefont {Rosenbusch}}]{Rosenbusch2012}%
  \BibitemOpen
  \bibfield  {author} {\bibinfo {author} {\bibfnamefont {W.}~\bibnamefont
  {Maineult}}, \bibinfo {author} {\bibfnamefont {C.}~\bibnamefont {Deutsch}},
  \bibinfo {author} {\bibfnamefont {K.}~\bibnamefont {Gibble}}, \bibinfo
  {author} {\bibfnamefont {J.}~\bibnamefont {Reichel}}, \ and\ \bibinfo
  {author} {\bibfnamefont {P.}~\bibnamefont {Rosenbusch}},\ }\href@noop {}
  {\bibfield  {journal} {\bibinfo  {journal} {Phys. Rev. Lett.}\ }\textbf
  {\bibinfo {volume} {109}},\ \bibinfo {pages} {020407} (\bibinfo {year}
  {2012})}\BibitemShut {NoStop}%
\bibitem [{\citenamefont {Lemke}\ \emph {et~al.}(2009)\citenamefont {Lemke},
  \citenamefont {Ludlow}, \citenamefont {Barber}, \citenamefont {Fortier},
  \citenamefont {Diddams}, \citenamefont {Jiang}, \citenamefont {Jefferts},
  \citenamefont {Heavner}, \citenamefont {Parker},\ and\ \citenamefont
  {Oates}}]{Oates2009}%
  \BibitemOpen
  \bibfield  {author} {\bibinfo {author} {\bibfnamefont {N.~D.}\ \bibnamefont
  {Lemke}}, \bibinfo {author} {\bibfnamefont {A.~D.}\ \bibnamefont {Ludlow}},
  \bibinfo {author} {\bibfnamefont {Z.~W.}\ \bibnamefont {Barber}}, \bibinfo
  {author} {\bibfnamefont {T.~M.}\ \bibnamefont {Fortier}}, \bibinfo {author}
  {\bibfnamefont {S.~A.}\ \bibnamefont {Diddams}}, \bibinfo {author}
  {\bibfnamefont {Y.}~\bibnamefont {Jiang}}, \bibinfo {author} {\bibfnamefont
  {S.~R.}\ \bibnamefont {Jefferts}}, \bibinfo {author} {\bibfnamefont {T.~P.}\
  \bibnamefont {Heavner}}, \bibinfo {author} {\bibfnamefont {T.~E.}\
  \bibnamefont {Parker}}, \ and\ \bibinfo {author} {\bibfnamefont {C.~W.}\
  \bibnamefont {Oates}},\ }\href@noop {} {\bibfield  {journal} {\bibinfo
  {journal} {Phys. Rev. Lett.}\ }\textbf {\bibinfo {volume} {103}},\ \bibinfo
  {pages} {063001} (\bibinfo {year} {2009})}\BibitemShut {NoStop}%
\bibitem [{\citenamefont {Campbell}\ \emph {et~al.}(2009)\citenamefont
  {Campbell}, \citenamefont {Boyd}, \citenamefont {Thomsen}, \citenamefont
  {Martin}, \citenamefont {Blatt}, \citenamefont {Swallows}, \citenamefont
  {Nicholson}, \citenamefont {Fortier}, \citenamefont {Oates}, \citenamefont
  {Diddams}, \citenamefont {Lemke}, \citenamefont {Naidon}, \citenamefont
  {Julienne}, \citenamefont {Ye},\ and\ \citenamefont {Ludlow}}]{Ludlow2009}%
  \BibitemOpen
  \bibfield  {author} {\bibinfo {author} {\bibfnamefont {G.~K.}\ \bibnamefont
  {Campbell}}, \bibinfo {author} {\bibfnamefont {M.~M.}\ \bibnamefont {Boyd}},
  \bibinfo {author} {\bibfnamefont {J.~W.}\ \bibnamefont {Thomsen}}, \bibinfo
  {author} {\bibfnamefont {M.~J.}\ \bibnamefont {Martin}}, \bibinfo {author}
  {\bibfnamefont {S.}~\bibnamefont {Blatt}}, \bibinfo {author} {\bibfnamefont
  {M.~D.}\ \bibnamefont {Swallows}}, \bibinfo {author} {\bibfnamefont {T.~L.}\
  \bibnamefont {Nicholson}}, \bibinfo {author} {\bibfnamefont {T.}~\bibnamefont
  {Fortier}}, \bibinfo {author} {\bibfnamefont {C.~W.}\ \bibnamefont {Oates}},
  \bibinfo {author} {\bibfnamefont {S.~A.}\ \bibnamefont {Diddams}}, \bibinfo
  {author} {\bibfnamefont {N.~D.}\ \bibnamefont {Lemke}}, \bibinfo {author}
  {\bibfnamefont {P.}~\bibnamefont {Naidon}}, \bibinfo {author} {\bibfnamefont
  {P.}~\bibnamefont {Julienne}}, \bibinfo {author} {\bibfnamefont
  {J.}~\bibnamefont {Ye}}, \ and\ \bibinfo {author} {\bibfnamefont {A.~D.}\
  \bibnamefont {Ludlow}},\ }\href@noop {} {\bibfield  {journal} {\bibinfo
  {journal} {Science}\ }\textbf {\bibinfo {volume} {324}},\ \bibinfo {pages}
  {360} (\bibinfo {year} {2009})}\BibitemShut {NoStop}%
\bibitem [{\citenamefont {Blatt}\ \emph {et~al.}(2009)\citenamefont {Blatt},
  \citenamefont {Thomsen}, \citenamefont {Campbell}, \citenamefont {Ludlow},
  \citenamefont {Swallows}, \citenamefont {Martin}, \citenamefont {Boyd},\ and\
  \citenamefont {Ye}}]{Ye2009}%
  \BibitemOpen
  \bibfield  {author} {\bibinfo {author} {\bibfnamefont {S.}~\bibnamefont
  {Blatt}}, \bibinfo {author} {\bibfnamefont {J.~W.}\ \bibnamefont {Thomsen}},
  \bibinfo {author} {\bibfnamefont {G.~K.}\ \bibnamefont {Campbell}}, \bibinfo
  {author} {\bibfnamefont {A.~D.}\ \bibnamefont {Ludlow}}, \bibinfo {author}
  {\bibfnamefont {M.~D.}\ \bibnamefont {Swallows}}, \bibinfo {author}
  {\bibfnamefont {M.~J.}\ \bibnamefont {Martin}}, \bibinfo {author}
  {\bibfnamefont {M.~M.}\ \bibnamefont {Boyd}}, \ and\ \bibinfo {author}
  {\bibfnamefont {J.}~\bibnamefont {Ye}},\ }\href@noop {} {\bibfield  {journal}
  {\bibinfo  {journal} {Phys. Rev. A}\ }\textbf {\bibinfo {volume} {80}},\
  \bibinfo {pages} {052703} (\bibinfo {year} {2009})}\BibitemShut {NoStop}%
\bibitem [{\citenamefont {Swallows}\ \emph {et~al.}(2011)\citenamefont
  {Swallows}, \citenamefont {Bishof}, \citenamefont {Lin}, \citenamefont
  {Blatt}, \citenamefont {Martin}, \citenamefont {Rey},\ and\ \citenamefont
  {Ye}}]{Ye2011A}%
  \BibitemOpen
  \bibfield  {author} {\bibinfo {author} {\bibfnamefont {M.~D.}\ \bibnamefont
  {Swallows}}, \bibinfo {author} {\bibfnamefont {M.}~\bibnamefont {Bishof}},
  \bibinfo {author} {\bibfnamefont {Y.}~\bibnamefont {Lin}}, \bibinfo {author}
  {\bibfnamefont {S.}~\bibnamefont {Blatt}}, \bibinfo {author} {\bibfnamefont
  {M.~J.}\ \bibnamefont {Martin}}, \bibinfo {author} {\bibfnamefont {A.~M.}\
  \bibnamefont {Rey}}, \ and\ \bibinfo {author} {\bibfnamefont
  {J.}~\bibnamefont {Ye}},\ }\href@noop {} {\bibfield  {journal} {\bibinfo
  {journal} {Science}\ }\textbf {\bibinfo {volume} {331}},\ \bibinfo {pages}
  {1043} (\bibinfo {year} {2011})}\BibitemShut {NoStop}%
\bibitem [{\citenamefont {Bishof}\ \emph {et~al.}(2011)\citenamefont {Bishof},
  \citenamefont {Lin}, \citenamefont {Swallows}, \citenamefont {Gorshkov},
  \citenamefont {Ye},\ and\ \citenamefont {Rey}}]{Rey2011A}%
  \BibitemOpen
  \bibfield  {author} {\bibinfo {author} {\bibfnamefont {M.}~\bibnamefont
  {Bishof}}, \bibinfo {author} {\bibfnamefont {Y.}~\bibnamefont {Lin}},
  \bibinfo {author} {\bibfnamefont {M.~D.}\ \bibnamefont {Swallows}}, \bibinfo
  {author} {\bibfnamefont {A.~V.}\ \bibnamefont {Gorshkov}}, \bibinfo {author}
  {\bibfnamefont {J.}~\bibnamefont {Ye}}, \ and\ \bibinfo {author}
  {\bibfnamefont {A.~M.}\ \bibnamefont {Rey}},\ }\href@noop {} {\bibfield
  {journal} {\bibinfo  {journal} {Phys. Rev. Lett.}\ }\textbf {\bibinfo
  {volume} {106}},\ \bibinfo {pages} {250801} (\bibinfo {year}
  {2011})}\BibitemShut {NoStop}%
\bibitem [{\citenamefont {Lemke}\ \emph {et~al.}(2011)\citenamefont {Lemke},
  \citenamefont {von Stecher}, \citenamefont {Sherman}, \citenamefont {Rey},
  \citenamefont {Oates},\ and\ \citenamefont {Ludlow}}]{Ludlow2011}%
  \BibitemOpen
  \bibfield  {author} {\bibinfo {author} {\bibfnamefont {N.~D.}\ \bibnamefont
  {Lemke}}, \bibinfo {author} {\bibfnamefont {J.}~\bibnamefont {von Stecher}},
  \bibinfo {author} {\bibfnamefont {J.~A.}\ \bibnamefont {Sherman}}, \bibinfo
  {author} {\bibfnamefont {A.~M.}\ \bibnamefont {Rey}}, \bibinfo {author}
  {\bibfnamefont {C.~W.}\ \bibnamefont {Oates}}, \ and\ \bibinfo {author}
  {\bibfnamefont {A.~D.}\ \bibnamefont {Ludlow}},\ }\href@noop {} {\bibfield
  {journal} {\bibinfo  {journal} {Phys. Rev. Lett.}\ }\textbf {\bibinfo
  {volume} {107}},\ \bibinfo {pages} {103902} (\bibinfo {year}
  {2011})}\BibitemShut {NoStop}%
\bibitem [{\citenamefont {Bartenstein}\ \emph {et~al.}(2005)\citenamefont
  {Bartenstein} \emph {et~al.}}]{Julienne05}%
  \BibitemOpen
  \bibfield  {author} {\bibinfo {author} {\bibfnamefont {M.}~\bibnamefont
  {Bartenstein}} \emph {et~al.},\ }\href@noop {} {\bibfield  {journal}
  {\bibinfo  {journal} {Phys. Rev. Lett.}\ }\textbf {\bibinfo {volume} {94}},\
  \bibinfo {pages} {103201} (\bibinfo {year} {2005})}\BibitemShut {NoStop}%
\bibitem [{Jul()}]{JuliennePrivate}%
  \BibitemOpen
  \href@noop {} {}\bibinfo {note} {P. Julienne (private
  communication).}\BibitemShut {Stop}%
\bibitem [{\citenamefont {Du}\ \emph {et~al.}(2008)\citenamefont {Du},
  \citenamefont {Luo}, \citenamefont {Clancy},\ and\ \citenamefont
  {Thomas}}]{Thomas2008}%
  \BibitemOpen
  \bibfield  {author} {\bibinfo {author} {\bibfnamefont {X.}~\bibnamefont
  {Du}}, \bibinfo {author} {\bibfnamefont {L.}~\bibnamefont {Luo}}, \bibinfo
  {author} {\bibfnamefont {B.}~\bibnamefont {Clancy}}, \ and\ \bibinfo {author}
  {\bibfnamefont {J.~E.}\ \bibnamefont {Thomas}},\ }\href@noop {} {\bibfield
  {journal} {\bibinfo  {journal} {Phys. Rev. Lett.}\ }\textbf {\bibinfo
  {volume} {101}},\ \bibinfo {pages} {150401} (\bibinfo {year}
  {2008})}\BibitemShut {NoStop}%
\bibitem [{Note1()}]{Note1}%
  \BibitemOpen
  \bibinfo {note} {For each $\theta _1$, $\theta _2$, we adjust the
  interrogation time $T$ to keep a constant Ramsey fringe linewidth, with an
  effective interrogation time of $2.7 \protect \tmspace +\thinmuskip {.1667em}
  {\protect \mathrm {ms}}$. The Raman $\pi /2$ pulse duration is $450 \protect
  \tmspace +\thinmuskip {.1667em} \mu {\protect \mathrm {s}}$. To always
  resolve the trap sidebands, we change the pulse area for $\theta > \pi /2$
  ($\theta < \pi /2$) by increasing (decreasing) the duration (amplitude) of
  the Raman beam pulse and use a smooth turn-on and turn-off, which are $37.5
  \protect \tmspace +\thinmuskip {.1667em} \mu {\protect \mathrm {s}}$ long.
  For $\theta < \pi /2$, the total optical intensity is maintained by adding an
  optical field off-resonant with the two-photon Raman transition.}\BibitemShut
  {Stop}%
\bibitem [{Note2()}]{Note2}%
  \BibitemOpen
  \bibinfo {note} {We consider a contact interaction $V({\protect \bf
  {r}}_1,{\protect \bf {r}}_2) = \protect \frac {4 \pi \hbar ^2 a_{\delimiter
  "3223379 \delimiter "3222378 }}{m} \protect \tmspace +\thinmuskip {.1667em}
  \delta ({\protect \bf {r}}_1 - {\protect \bf {r}}_2)$. Here, $g = \protect
  \frac {2 h a_{\delimiter "3223379 \delimiter "3222378 }}{m} \DOTSI \intop
  \ilimits@ \psi _\alpha ^2({\protect \bf {r}}) \protect \tmspace +\thinmuskip
  {.1667em} \psi _\beta ^2({\protect \bf {r}}) \protect \tmspace +\thinmuskip
  {.1667em} dV$.}\BibitemShut {Stop}%
\bibitem [{Note3()}]{Note3}%
  \BibitemOpen
  \bibinfo {note} {We use scattering lengths much smaller than the deBroglie
  wavelength, $2000 a_0$, so that trap state changing (lateral) collisions can
  be neglected.}\BibitemShut {Stop}%
\bibitem [{Note4()}]{Note4}%
  \BibitemOpen
  \bibinfo {note} {We define $\Delta \Omega ^2 \equiv {\delimiter "426830A }
  g_{\alpha \beta } \Delta \Omega _{\alpha \beta }^2 {\delimiter "526930B }
  /{\delimiter "426830A } g_{\alpha \beta } {\delimiter "526930B }$, where
  ${\delimiter "426830A } \protect \tmspace +\thinmuskip {.1667em} {\delimiter
  "526930B }$ denotes an ensemble average over all atom pairs in states $\psi
  _\alpha $ and $\psi _\beta $. Then, for small $\Delta \Omega $ and $\Omega $,
  $\Delta \nu $ is $ (g/\pi ) \protect \tmspace +\thinmuskip {.1667em} (\Delta
  \Omega /\Omega )^2$ for two atoms and $( (N - 1) {\delimiter "426830A }
  g_{\alpha \beta }{\delimiter "526930B }/\pi ) \protect \tmspace +\thinmuskip
  {.1667em} ( \Delta \Omega /\Omega )^2$ for $N$ atoms. Here, $\Omega $ is the
  average Rabi frequency. In the context of the Rabi flopping in Fig.~\ref
  {Fig3}(b), a reasonable definition of $\Delta \Omega ^2$ is simply
  ${\delimiter "426830A } \Delta \Omega _{\alpha \beta }^2 {\delimiter "526930B
  }$, but this neglects important correlations between $\Delta \Omega _{\alpha
  \beta }^2$ and ${g}_{\alpha \beta }$.}\BibitemShut {Stop}%
\bibitem [{\citenamefont {Harber}\ \emph {et~al.}(2002)\citenamefont {Harber},
  \citenamefont {Lewandowski}, \citenamefont {McGuirk},\ and\ \citenamefont
  {Cornell}}]{Cornell2002}%
  \BibitemOpen
  \bibfield  {author} {\bibinfo {author} {\bibfnamefont {D.~M.}\ \bibnamefont
  {Harber}}, \bibinfo {author} {\bibfnamefont {H.~J.}\ \bibnamefont
  {Lewandowski}}, \bibinfo {author} {\bibfnamefont {J.~M.}\ \bibnamefont
  {McGuirk}}, \ and\ \bibinfo {author} {\bibfnamefont {E.~A.}\ \bibnamefont
  {Cornell}},\ }\href@noop {} {\bibfield  {journal} {\bibinfo  {journal} {Phys.
  Rev. A}\ }\textbf {\bibinfo {volume} {66}},\ \bibinfo {pages} {053616}
  (\bibinfo {year} {2002})}\BibitemShut {NoStop}%
\bibitem [{Note5()}]{Note5}%
  \BibitemOpen
  \bibinfo {note} {The Raman beams are detuned to the red (blue) of the D2 (D1)
  line by 4.8 GHz (5.2 GHz) to minimize spontaneous emission and the
  differential light shift $\Delta \nu _{\protect \mathrm {LS}}$. For a trap
  state $\psi _\alpha $, $\Delta \nu _{ \protect \mathrm {LS}}$ is $ \simeq
  0.34 \protect \tmspace +\thinmuskip {.1667em} \Omega _\alpha /(2 \pi )$. Weak
  laser beams clear the small fraction of atoms in states populated by
  spontaneous emission to eliminate extraneous collisional shifts.}\BibitemShut
  {Stop}%
\bibitem [{\citenamefont {Regal}\ and\ \citenamefont {Jin}(2003)}]{Jin2003B}%
  \BibitemOpen
  \bibfield  {author} {\bibinfo {author} {\bibfnamefont {C.~A.}\ \bibnamefont
  {Regal}}\ and\ \bibinfo {author} {\bibfnamefont {D.~S.}\ \bibnamefont
  {Jin}},\ }\href@noop {} {\bibfield  {journal} {\bibinfo  {journal} {Phys.
  Rev. Lett.}\ }\textbf {\bibinfo {volume} {90}},\ \bibinfo {pages} {230404}
  (\bibinfo {year} {2003})}\BibitemShut {NoStop}%
\end{thebibliography}
\end{document}